\pdfoutput=1
\documentclass[12pt,a4paper]{article}

\usepackage{ifthen} 
\newboolean{pdflatex}
\setboolean{pdflatex}{true} 

\newboolean{articletitles}
\setboolean{articletitles}{true} 

\newboolean{uprightparticles}
\setboolean{uprightparticles}{false} 

\newboolean{inbibliography}
\setboolean{inbibliography}{false} 

\def\paperauthors{LHCb collaboration} 

\newcommand{\mevnospace}{\ensuremath{\mathrm{Me\kern -0.1em V}}\xspace}

\newcommand{\pislow}{{\ensuremath{\pi_{s}^{\pm}}}\xspace}

\newcommand{\sigmabstarstar}{{\ensuremath{\Sigmares_b{(6097)}}}\xspace}
\newcommand{\sigmabstarstarpm}{{\ensuremath{\Sigmares_b{(6097)}^\pm}}\xspace}
\newcommand{\sigmabstarstarp}{{\ensuremath{\Sigmares_b{(6097)}^{+}}}\xspace}
\newcommand{\sigmabstarstarm}{{\ensuremath{\Sigmares_b{(6097)}^{-}}}\xspace}
\newcommand{\Qformula}{{\ensuremath{Q \equiv m(\Lb \pi^\pm) - m(\Lb) - m(\pi^\pm)}}\xspace}

\newcommand{\lambdabpip}{{\ensuremath{\Lb \pi^+}}\xspace}
\newcommand{\lambdabpim}{{\ensuremath{\Lb \pi^-}}\xspace}
\newcommand{\lambdabpipm}{{\ensuremath{\Lb \pi^\pm}}\xspace}
\newcommand{\sigmab}{{\ensuremath{\Sigmares_b}}\xspace}
\newcommand{\sigmabstar}{{\ensuremath{\Sigmares_b^{*}}}\xspace}
\newcommand{\sigmabpm}{{\ensuremath{\Sigmares_b^\pm}}\xspace}
\newcommand{\sigmabp}{{\ensuremath{\Sigmares_b^+}}\xspace}
\newcommand{\sigmabm}{{\ensuremath{\Sigmares_b^-}}\xspace}
\newcommand{\sigmabstarpm}{{\ensuremath{\Sigmares_b^{*\pm}}}\xspace}
\newcommand{\sigmabstarp}{{\ensuremath{\Sigmares_b^{*+}}}\xspace}
\newcommand{\sigmabstarm}{{\ensuremath{\Sigmares_b^{*-}}}\xspace}
\newcommand{\onecolumngrid}{}
\newcommand{\twocolumngrid}{}

\def\paperauthors{LHCb collaboration} 

\def\paperasciititle{Observation of two resonances in the lambdab0pi system and precise measurement of Sigmab and Sigmabstar properties} 
\def\papertitle{Observation of two resonances in the \lambdabpipm systems and precise measurement of \sigmabpm and \sigmabstarpm properties} 
\def\paperkeywords{{High Energy Physics}, {LHCb}} 
\def\papercopyright{\the\year\ CERN for the benefit of the LHCb collaboration} 
\def\paperlicence{CC-BY-4.0 licence}
\def\paperlicenceurl{https://creativecommons.org/licenses/by/4.0/}

\usepackage[top=1in, bottom=1.25in, left=1in, right=1in]{geometry}

\usepackage{microtype}
\usepackage{lineno}  
\usepackage{xspace} 
\usepackage{caption} 

\usepackage{graphicx}  
\usepackage{color}
\usepackage{colortbl}
\graphicspath{{./figs/}} 

\usepackage{amsmath} 
\usepackage{amssymb}
\usepackage{amsfonts}
\usepackage{upgreek} 

\newcommand*\patchAmsMathEnvironmentForLineno[1]{%
\expandafter\let\csname old#1\expandafter\endcsname\csname #1\endcsname
\expandafter\let\csname oldend#1\expandafter\endcsname\csname
end#1\endcsname
 \renewenvironment{#1}%
   {\linenomath\csname old#1\endcsname}%
   {\csname oldend#1\endcsname\endlinenomath}%
}
\newcommand*\patchBothAmsMathEnvironmentsForLineno[1]{%
  \patchAmsMathEnvironmentForLineno{#1}%
  \patchAmsMathEnvironmentForLineno{#1*}%
}
\AtBeginDocument{%
\patchBothAmsMathEnvironmentsForLineno{equation}%
\patchBothAmsMathEnvironmentsForLineno{align}%
\patchBothAmsMathEnvironmentsForLineno{flalign}%
\patchBothAmsMathEnvironmentsForLineno{alignat}%
\patchBothAmsMathEnvironmentsForLineno{gather}%
\patchBothAmsMathEnvironmentsForLineno{multline}%
\patchBothAmsMathEnvironmentsForLineno{eqnarray}%
}

\usepackage{hyperxmp}

\usepackage[pdftex,
            pdfauthor={\paperauthors},
            pdftitle={\paperasciititle},
            pdfkeywords={\paperkeywords},
            pdfcopyright={Copyright (C) \papercopyright},
            pdflicenseurl={\paperlicenceurl}]{hyperref}

\usepackage[all]{hypcap} 

\usepackage{xspace} 
\usepackage{upgreek}

\def\lhcb   {\mbox{LHCb}\xspace}

\def\cdf    {\mbox{CDF}\xspace}

\def\MagUp {\mbox{\em Mag\kern -0.05em Up}\xspace}

\ifthenelse{\boolean{uprightparticles}}%
{

 \def\Ppsi        {\ensuremath{\uppsi}\xspace}

 \def\PDelta      {\ensuremath{\Delta}\xspace}                 
 \def\PXi      {\ensuremath{\Xi}\xspace}                 
 \def\PLambda      {\ensuremath{\Lambda}\xspace}                 
 \def\PSigma      {\ensuremath{\Sigma}\xspace}                 
 \def\POmega      {\ensuremath{\Omega}\xspace}                 
 \def\PUpsilon      {\ensuremath{\Upsilon}\xspace}                 
 
 \def\PB      {\ensuremath{\mathrm{B}}\xspace}                 
                  
 \def\PD      {\ensuremath{\mathrm{D}}\xspace}

 \def\PJ      {\ensuremath{\mathrm{J}}\xspace}                 
 \def\PK      {\ensuremath{\mathrm{K}}\xspace}

 \def\Pb      {\ensuremath{\mathrm{b}}\xspace}                 
 \def\Pc      {\ensuremath{\mathrm{c}}\xspace}

 \def\Pi      {\ensuremath{\mathrm{i}}\xspace}

}
{

 \def\Ppsi        {\ensuremath{\psi}\xspace}                 
                  
 \mathchardef\PDelta="7101
 \mathchardef\PXi="7104
 \mathchardef\PLambda="7103
 \mathchardef\PSigma="7106
 \mathchardef\POmega="710A
 \mathchardef\PUpsilon="7107
                  
 \def\PB      {\ensuremath{B}\xspace}                 
                  
 \def\PD      {\ensuremath{D}\xspace}

 \def\PJ      {\ensuremath{J}\xspace}                 
 \def\PK      {\ensuremath{K}\xspace}

 \def\Pb      {\ensuremath{b}\xspace}                 
 \def\Pc      {\ensuremath{c}\xspace}

 \def\Pi      {\ensuremath{i}\xspace}

}

\makeatletter
\ifcase \@ptsize \relax
  \newcommand{\miniscule}{\@setfontsize\miniscule{4}{5}}
\or
  \newcommand{\miniscule}{\@setfontsize\miniscule{5}{6}}
\or
  \newcommand{\miniscule}{\@setfontsize\miniscule{5}{6}}
\fi
\makeatother

\DeclareRobustCommand{\optbar}[1]{\shortstack{{\miniscule (\rule[.5ex]{1.25em}{.18mm})}
  \\ [-.7ex] $#1$}}

\def\cquark    {{\ensuremath{\Pc}}\xspace}

\def\bquark    {{\ensuremath{\Pb}}\xspace}

\def\kaon    {{\ensuremath{\PK}}\xspace}
  \def\Kbar    {{\kern 0.2em\overline{\kern -0.2em \PK}{}}\xspace}

\def\KorKbar    {\kern 0.18em\optbar{\kern -0.18em K}{}\xspace}

\def\KS      {{\ensuremath{\kaon^0_{\mathrm{ \scriptscriptstyle S}}}}\xspace}

  \def\Dbar    {{\kern 0.2em\overline{\kern -0.2em \PD}{}}\xspace}
\def\D       {{\ensuremath{\PD}}\xspace}

\def\DorDbar    {\kern 0.18em\optbar{\kern -0.18em D}{}\xspace}
\def\Dz      {{\ensuremath{\D^0}}\xspace}

\def\Dstarp  {{\ensuremath{\D^{*+}}}\xspace}

\def\B       {{\ensuremath{\PB}}\xspace}
\def\Bbar    {{\ensuremath{\kern 0.18em\overline{\kern -0.18em \PB}{}}}\xspace}

\def\BorBbar    {\kern 0.18em\optbar{\kern -0.18em B}{}\xspace}

\def\Bu      {{\ensuremath{\B^+}}\xspace}

\def\jpsi     {{\ensuremath{{\PJ\mskip -3mu/\mskip -2mu\Ppsi\mskip 2mu}}}\xspace}

  \def\Y#1S{\ensuremath{\PUpsilon{(#1S)}}\xspace}

\def\Lz          {{\ensuremath{\PLambda}}\xspace}
\def\Lbar        {{\ensuremath{\kern 0.1em\overline{\kern -0.1em\PLambda}}}\xspace}
\def\LorLbar    {\kern 0.18em\optbar{\kern -0.18em \PLambda}{}\xspace}

\def\Sigmares    {{\ensuremath{\PSigma}}\xspace}
\def\Sigmaresbar {{\ensuremath{\overline \Sigmares}}\xspace}

\def\Lb      {{\ensuremath{\Lz^0_\bquark}}\xspace}
\def\Lbbar   {{\ensuremath{\Lbar{}^0_\bquark}}\xspace}
\def\Lc      {{\ensuremath{\Lz^+_\cquark}}\xspace}

\def\AT#1     {\ensuremath{A_{\mathrm{T}}^{#1}}\xspace}           

\def\C#1      {\ensuremath{\mathcal{C}_{#1}}\xspace}                       
\def\Cp#1     {\ensuremath{\mathcal{C}_{#1}^{'}}\xspace}                    
\def\Ceff#1   {\ensuremath{\mathcal{C}_{#1}^{\mathrm{(eff)}}}\xspace}        
\def\Cpeff#1  {\ensuremath{\mathcal{C}_{#1}^{'\mathrm{(eff)}}}\xspace}       
\def\Ope#1    {\ensuremath{\mathcal{O}_{#1}}\xspace}                       
\def\Opep#1   {\ensuremath{\mathcal{O}_{#1}^{'}}\xspace}                    

\newcommand{\tev}{\ifthenelse{\boolean{inbibliography}}{\ensuremath{~T\kern -0.05em eV}}{\ensuremath{\mathrm{\,Te\kern -0.1em V}}}\xspace}
\newcommand{\Tev}{\ifthenelse{\boolean{inbibliography}}{\ensuremath{~T\kern -0.05em eV}}{\ensuremath{\mathrm{\,Te\kern -0.1em V}}}\xspace}
\newcommand{\gev}{\ensuremath{\mathrm{\,Ge\kern -0.1em V}}\xspace}
\newcommand{\mev}{\ensuremath{\mathrm{\,Me\kern -0.1em V}}\xspace}
\newcommand{\kev}{\ensuremath{\mathrm{\,ke\kern -0.1em V}}\xspace}
\newcommand{\ev}{\ensuremath{\mathrm{\,e\kern -0.1em V}}\xspace}
\newcommand{\gevc}{\ensuremath{{\mathrm{\,Ge\kern -0.1em V\!/}c}}\xspace}
\newcommand{\mevc}{\ensuremath{{\mathrm{\,Me\kern -0.1em V\!/}c}}\xspace}
\newcommand{\gevcc}{\ensuremath{{\mathrm{\,Ge\kern -0.1em V\!/}c^2}}\xspace}
\newcommand{\gevgevcccc}{\ensuremath{{\mathrm{\,Ge\kern -0.1em V^2\!/}c^4}}\xspace}
\newcommand{\mevcc}{\ensuremath{{\mathrm{\,Me\kern -0.1em V\!/}c^2}}\xspace}

\def\mum  {\ensuremath{{\,\upmu\mathrm{m}}}\xspace}

\def\invfb   {\ensuremath{\mbox{\,fb}^{-1}}\xspace}

\newcommand{\chisq}{\ensuremath{\chi^2}\xspace}

\newcommand{\chisqip}{\ensuremath{\chi^2_{\text{IP}}}\xspace}

\def\gsim{{~\raise.15em\hbox{$>$}\kern-.85em
          \lower.35em\hbox{$\sim$}~}\xspace}
\def\lsim{{~\raise.15em\hbox{$<$}\kern-.85em
          \lower.35em\hbox{$\sim$}~}\xspace}

\def\pt         {\ensuremath{p_{\mathrm{ T}}}\xspace}
\def\ptot       {\ensuremath{p}\xspace}

\def\tell1  {TELL1\xspace}
\def\ukl1   {UKL1\xspace}

\newcommand{\eg}{\mbox{\itshape e.g.}\xspace}

\usepackage{cite} 
\usepackage{mciteplus}

\usepackage{longtable} 
\usepackage{rotating}

\begin{document}

\renewcommand{\thefootnote}{\fnsymbol{footnote}}
\setcounter{footnote}{1}


\begin{titlepage}
\pagenumbering{roman}

\vspace*{-1.5cm}
\centerline{\large EUROPEAN ORGANIZATION FOR NUCLEAR RESEARCH (CERN)}
\vspace*{0.5cm}
\noindent
\begin{tabular*}{\linewidth}{lc@{\extracolsep{\fill}}r@{\extracolsep{0pt}}}
\ifthenelse{\boolean{pdflatex}}
{\vspace*{-1.5cm}\mbox{\!\!\!\includegraphics[width=.14\textwidth]{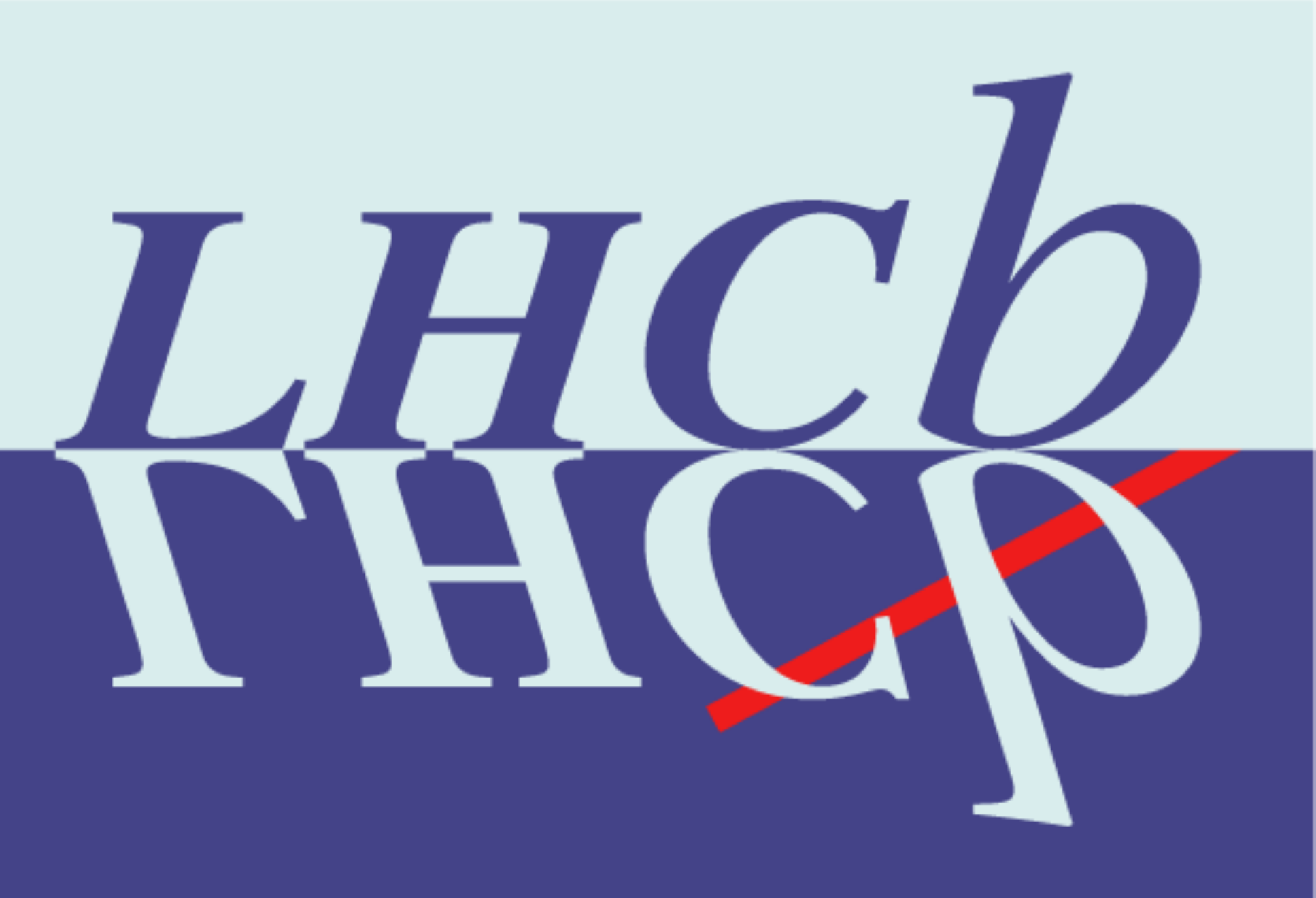}} & &}%
{\vspace*{-1.2cm}\mbox{\!\!\!\includegraphics[width=.12\textwidth]{lhcb-logo.eps}} & &}%
\\
 & & CERN-EP-2018-243 \\  
 & & LHCb-PAPER-2018-032 \\  
   & & \today \\ 
%
%
%
\end{tabular*}

\vspace*{2.0cm}

{\normalfont\bfseries\boldmath\huge
\begin{center}
  \papertitle 
\end{center}
}

\vspace*{1.0cm}

\begin{center}
\paperauthors\footnote{Authors are listed at the end of this Letter.}
\end{center}

\vspace{\fill}

\begin{abstract}
\noindent
The first observation of two structures consistent with resonances in the final states \lambdabpim and \lambdabpip
is reported using samples of $pp$ collision data collected by the LHCb experiment at $\sqrt{s} = 7$ and $8\Tev$,
corresponding to an integrated luminosity of 3\invfb.
The ground states \sigmabpm and \sigmabstarpm are also confirmed and their masses and widths are precisely measured.

\end{abstract}

\vspace*{1.0cm}

\begin{center}
  Submitted to Phys. Rev. Lett.
\end{center}

\vspace{\fill}

{\footnotesize 
\centerline{\copyright~\papercopyright, \href{\paperlicenceurl}{\paperlicence}.}}
\vspace*{2mm}

\end{titlepage}


\newpage
\setcounter{page}{2}
\mbox{~}
%
%
%
%

\cleardoublepage


\renewcommand{\thefootnote}{\arabic{footnote}}
\setcounter{footnote}{0}



\pagestyle{plain} 
\setcounter{page}{1}
\pagenumbering{arabic}


%

Bottom baryons are composed of a $b$ quark and two lighter quarks $(bqq')$.
In the constituent quark model~\cite{GELLMANN1964214,Zweig:352337}, such baryon states form multiplets according to the
symmetries of their flavor, spin, and spatial wave functions~\cite{RevModPhys.82.1095}.
The \Lb baryon is the lightest of the bottom baryons and forms an isospin ($I$) singlet $(bud)$ with spin-parity $J^P = {\frac{1}{2}}^+$.
Two $I=1$ triplets with $J^P = {\frac{1}{2}}^+$ (\sigmab) and $J^P = {\frac{3}{2}}^+$ (\sigmabstar) are expected,
with the spin of the flavor-symmetric $qq'$ diquark $S_{qq'} = 1$.
Four of those six states, the \sigmabpm and \sigmabstarpm baryons ($uub$ and $ddb$), have been observed
by the \cdf collaboration~\cite{Aaltonen:2007ar,CDF:2011ac}
and reported briefly in a previous \lhcb paper~\cite{LHCb-PAPER-2015-047}.
Beyond these ground states, radially and orbitally excited states are expected at higher masses, but only
a few excited baryons have been observed
in the bottom sector~\cite{LHCb-PAPER-2012-012,LHCb-PAPER-2018-013,PhysRevLett.108.252002,LHCb-PAPER-2014-061}.  
The search for and study of these states will cast light on the internal mechanisms governing the dynamics
of the constituent quarks~\cite{THAKKAR201757,PhysRevD.84.014025}.

In this Letter, we report the observation of structures in both the \lambdabpip and \lambdabpim mass distributions
(charge conjugation is implied throughout this article) 
using $pp$ collision data collected by the \lhcb experiment at $\sqrt{s} = 7$ and $8\Tev$,
corresponding to an integrated luminosity of 3\invfb.
We refer to these new states as \sigmabstarstarpm in the rest of the Letter.
We also measure precisely the masses and widths of the \sigmabpm and \sigmabstarpm ground states.
%
%
%
%
%

The \lhcb detector~\cite{Alves:2008zz,LHCb-DP-2014-002} is a single-arm forward
spectrometer covering the \mbox{pseudorapidity} range $2<\eta <5$,
designed for the study of particles containing \bquark or \cquark quarks.
The detector includes a high-precision tracking system
consisting of a silicon-strip vertex detector surrounding the $pp$
interaction region~\cite{LHCb-DP-2014-001}, a large-area silicon-strip detector located
upstream of a dipole magnet with a bending power of about
$4{\mathrm{\,Tm}}$, and three stations of silicon-strip detectors and straw
drift tubes~\cite{LHCb-DP-2013-003} placed downstream of the magnet.
The tracking system provides a measurement of the momentum, \ptot, of charged particles with
a relative uncertainty that varies from 0.5\% at low momentum to 1.0\% at 200\gev
(natural units with $c=\hbar=1$ are used throughout this Letter).
The momentum scale is calibrated using samples of $\jpsi \rightarrow \mu^+ \mu^-$
and $\Bu \rightarrow \jpsi K^+$ decays collected concurrently with the data sample
used for this analysis~\cite{LHCb-PAPER-2012-048,LHCb-PAPER-2013-011}.
The relative accuracy of this procedure is estimated to be
$3 \times 10^{-4}$ using samples of other fully reconstructed
$\bquark$-hadron, $\KS$, and narrow $\PUpsilon$ resonance decays.
The minimum distance of a track to a primary vertex (PV), the impact parameter (IP),
is measured with a resolution of $(15+29/\pt)\mum$,
where \pt is the component of the momentum transverse to the beam, in\,\gev.
Different types of charged hadrons are distinguished using information
from two ring-imaging Cherenkov detectors~\cite{LHCb-DP-2012-003}.
The online event selection is performed by a trigger~\cite{LHCb-DP-2012-004}
which consists of a hardware stage, based on information from the calorimeter and muon systems,
followed by a software stage, which applies a full event reconstruction.
The software trigger requires a two-, three- or four-track secondary vertex with significant displacement from all primary $pp$ interaction vertices. A multivariate algorithm~\cite{BBDT} 
is used for the identification of secondary vertices consistent with the decay of a $b$ hadron.
Simulated data samples are produced using the software packages described in
Refs.~\cite{Sjostrand:2006za,*Sjostrand:2007gs,LHCb-PROC-2010-056,Lange:2001uf,Golonka:2005pn,Allison:2006ve,*Agostinelli:2002hh}.
%
%
%
%
%

\begin{figure}[t]
\centering
\includegraphics[width=0.52\textwidth]{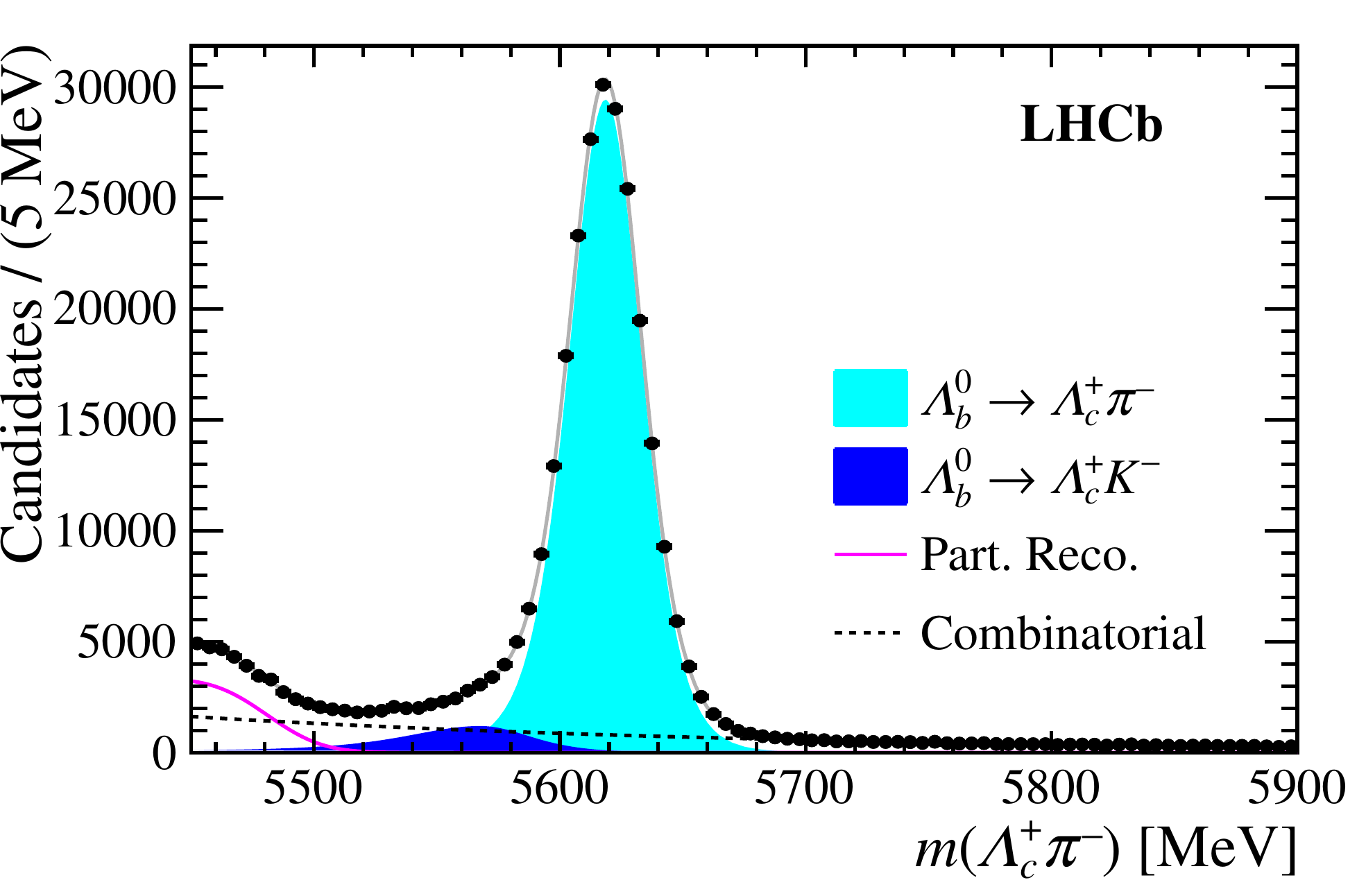}
\caption{\small
Mass distribution for the selected $\Lb \rightarrow \Lc \pi^-$ candidates.
The points show experimental data.
\label{fig:figure1}}
\end{figure}

Samples of \Lb candidates are formed from $\Lc \pi^-$ combinations,
where the \Lc baryon is reconstructed in the $p K^- \pi^+$ final state.
All charged particles used to form the $b$-hadron candidates are required to
have particle-identification information consistent with the appropriate mass hypothesis.
Misreconstructed tracks are suppressed by the use of a neural network trained
to discriminate between real and fake particles~\cite{DeCian:2255039}.
To suppress prompt background, all \Lb decay products are required to 
have significant \chisqip with respect to all PVs in the event,
where \chisqip is the difference in \chisq of the vertex fit of a given PV, when a particle is included or excluded from the fit.
The reconstructed \Lc vertex is required to have a good fit quality and to be significantly
displaced from all PVs in the event. The reconstructed \Lc mass must be within
a mass window of $\pm 25$\mev of its known value~\cite{PDG2016}.
Pion candidates that have large \chisqip with respect to all PVs
are combined with \Lc candidates to form \Lb candidates, requiring good vertex-fit quality
and separation of the \Lb decay point from any PV in the event.
A Boosted Decision Tree (BDT) discriminant~\cite{ROE2005577,AdaBoost} is used to further reduce the background.
The BDT exploits nineteen topological variables, including the \chisqip and \pt values of all the particles in the decay chain,
the \chisq values of the \Lb and \Lc decay vertices, their flight-distance significance,
and the angle between their momentum and direction of flight, defined by their production and decay vertices.
The BDT is trained using simulated \Lb signal decays and \Lb candidates in data in the sideband $5800 < m(\Lb) < 6000\mev$.
The signal candidates are refitted constraining the mass of the \Lc to its known value~\cite{PDG2016}
in order to improve the mass resolution~\cite{Hulsbergen:2005pu}.
The mass distribution of the selected $\Lb \rightarrow \Lc \pi^-$, $\Lc \rightarrow p K^- \pi^+$ candidates 
is shown in Fig.~\ref{fig:figure1}.
The mass spectrum is fitted with an asymmetric resolution function
for the signal component~\cite{LHCb-PAPER-2017-021},
plus a misreconstructed $\Lb \rightarrow \Lc K^-$ component whose yield is fixed relative to that of $\Lb \rightarrow \Lc \pi^-$,
an exponential function for the combinatorial background and an empirical function for partially reconstructed backgrounds
as described in Ref.~\cite{LHCb-PAPER-2017-021}.
The resulting \Lb signal yield is $234$,$270 \pm 900$. 

The \Lb candidates contained in a $\pm 50$\mev window around the peak maximum are then combined with a prompt pion,
hereafter referred to as \pislow, to form $\sigmabpm \rightarrow \Lb \pi^{\pm}$ combinations (along with
${\Sigmaresbar_b^\mp} \rightarrow \Lbbar \pi^{\mp}$).
Initially, $\pt(\pislow) > 200\mev$ and $\Qformula < 200\mev$ are required, where 
the \lambdabpipm mass is recomputed constraining the masses of the \Lc and \Lb baryons
to their known values~\cite{PDG2016}.
Then the search is extended to higher masses up to $Q = 600\mev$,
observing an additional peak in both \lambdabpim and \lambdabpip spectra.
A tighter transverse momentum cut $\pt(\pi_{s}^\pm) > 1000\mev$ is applied to remove the background from prompt pions.
%
%
%
%

The signal yields and parameters of the \sigmabpm, \sigmabstarpm and \sigmabstarstarpm resonances are determined
with extended unbinned maximum-likelihood fits to the $Q$-value distribution.
All signal components are modeled as relativistic Breit--Wigner functions~\cite{Jackson1964}
including Blatt--Weisskopf form factors~\cite{Blatt:1952ije} with a radius of $4\gev^{-1}$.
The orbital angular momentum $l$ between the \Lb baryon and \pislow candidate is taken to be 1 in all cases.
The relativistic Breit--Wigner functions are convolved with the detector resolution and corrected for a small distortion
in the shape induced by the \pt requirement on the \pislow meson.
The resolution models are determined from simulation, in which the three signal resonances are generated at the $Q$ values
found in data. The root-mean-square values of the resolution functions
for \sigmab, \sigmabstar, and \sigmabstarstar are
$0.99$, $1.13$ and $2.35$\mev, respectively,
all below the visible widths of the mass peaks and consistent with a resolution that scales as $\sqrt{Q}$.
Different empirical parameterizations are used for the two mass ranges.
For $0 < Q < 200$\mev the background shape is described by a smooth
threshold function~\cite{LHCb-PAPER-2012-030,LHCb-PAPER-2014-061,LHCb-PAPER-2017-002},
while for \mbox{$0 < Q < 600$\mev} a sigmoid function is used, as in Refs.~\cite{LHCb-PAPER-2013-028,LHCb-PAPER-2011-019}.
The background shapes are validated by using candidates in the data sidebands for a wide range of \pt requirements.
All of the masses, widths, and yields are free to vary in the fit, as are the background parameters; the resolutions of the signal components are fixed to the values found in simulation.
The fit models are validated with \mbox{pseudoexperiments} and no significant bias is found on any of the free parameters.
%
%
%
%

\begin{figure}[t]
\centering
\includegraphics[width=0.495\textwidth]{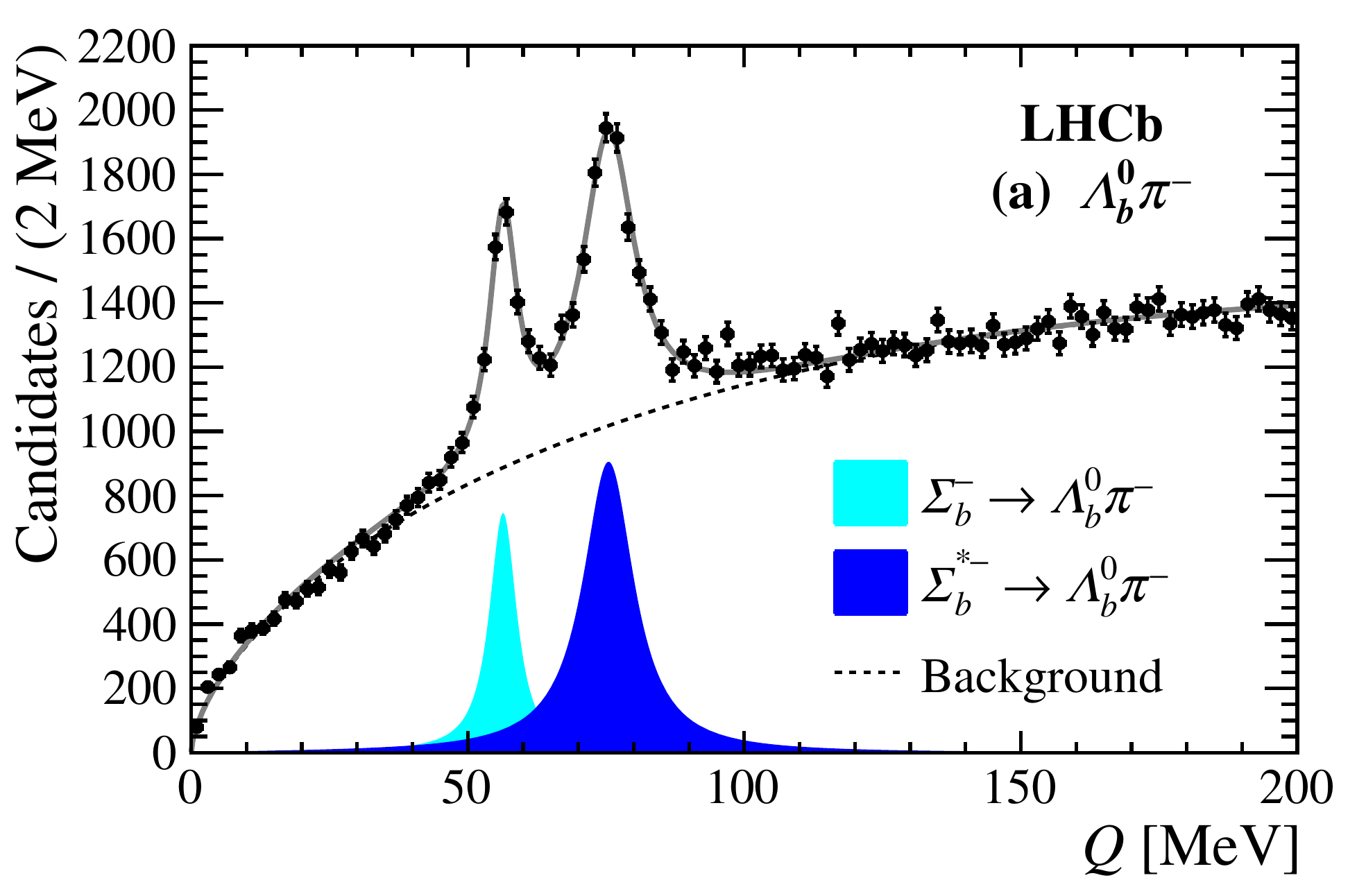}
\includegraphics[width=0.495\textwidth]{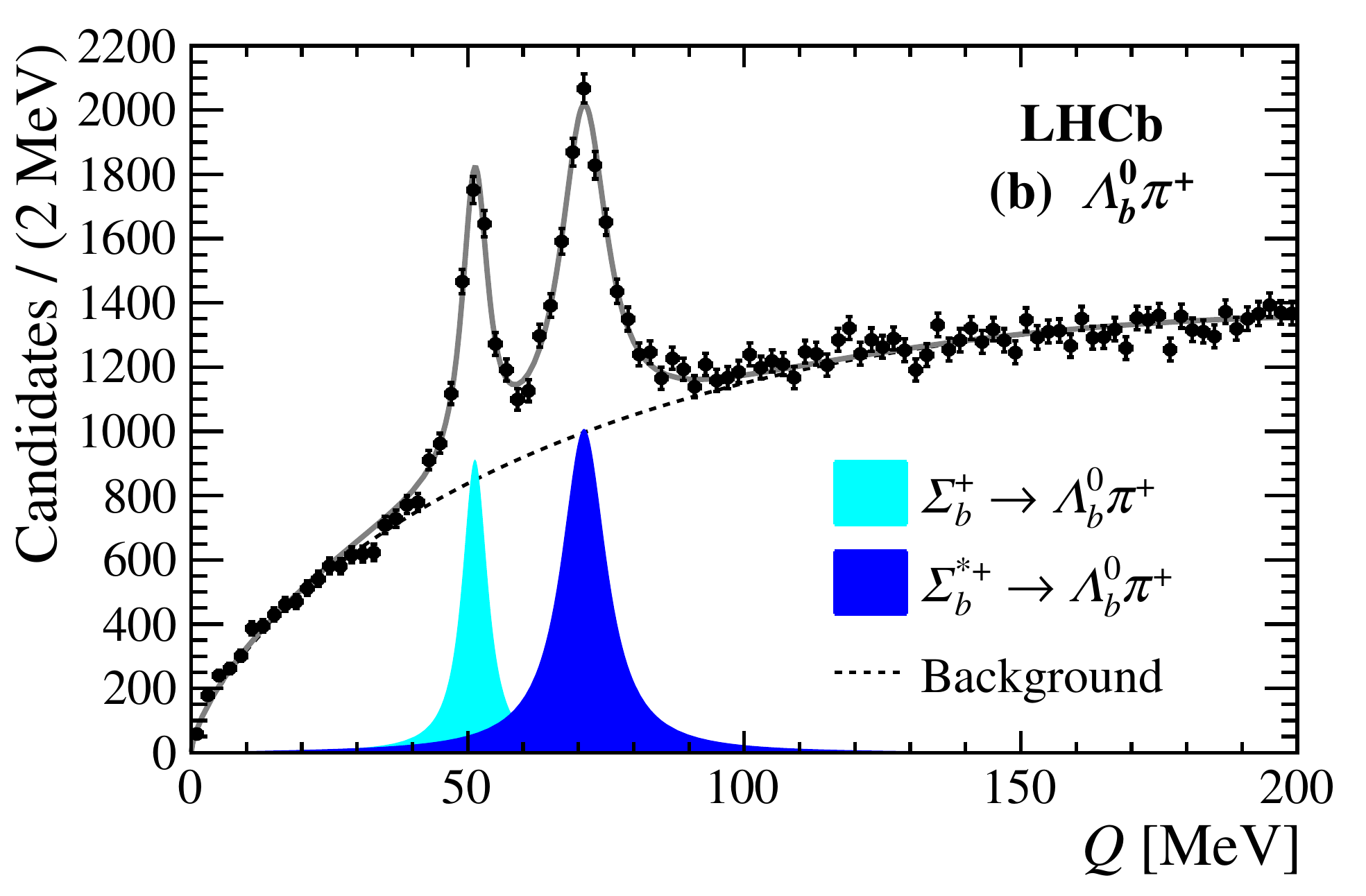}
\includegraphics[width=0.495\textwidth]{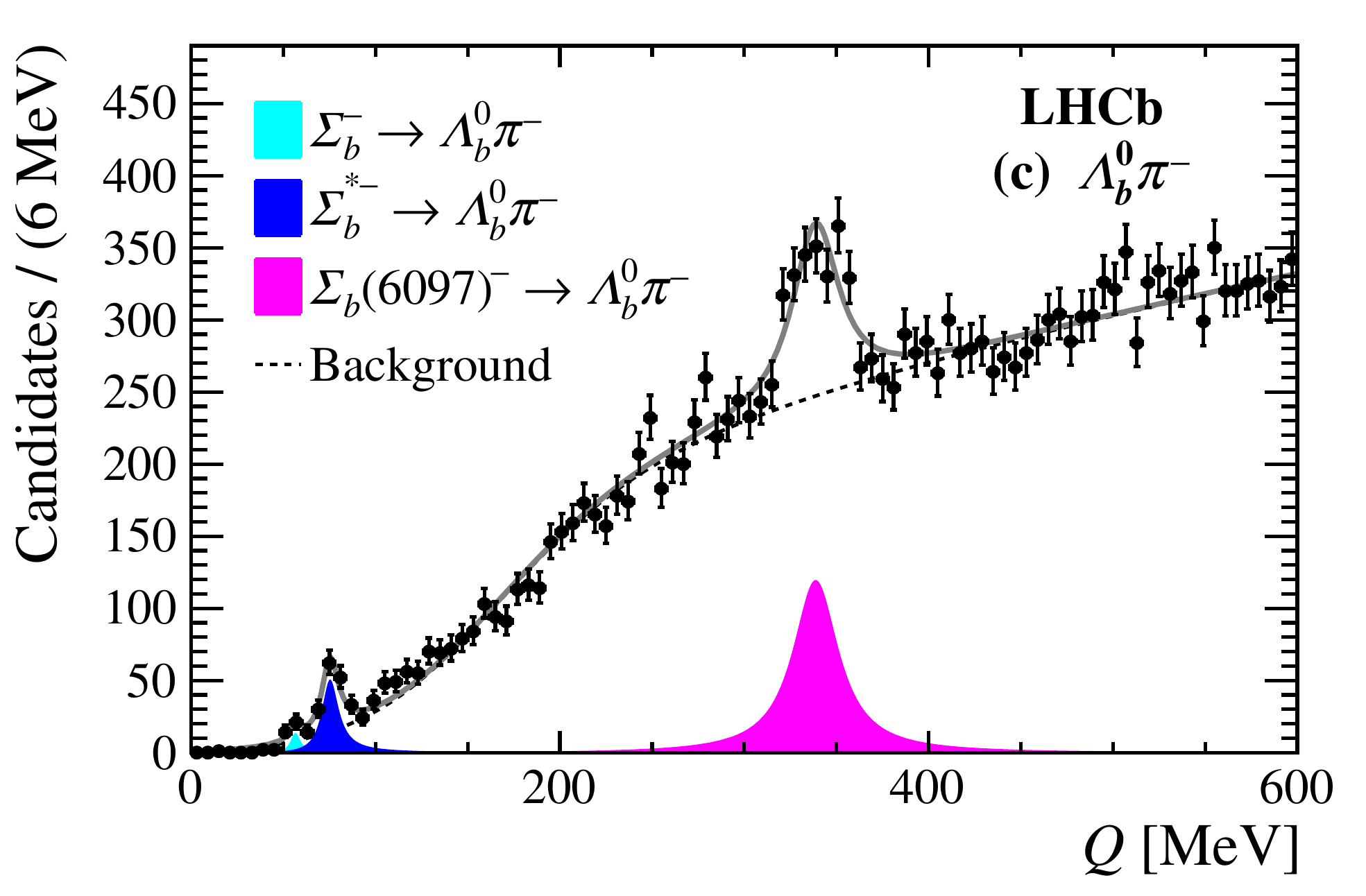}
\includegraphics[width=0.495\textwidth]{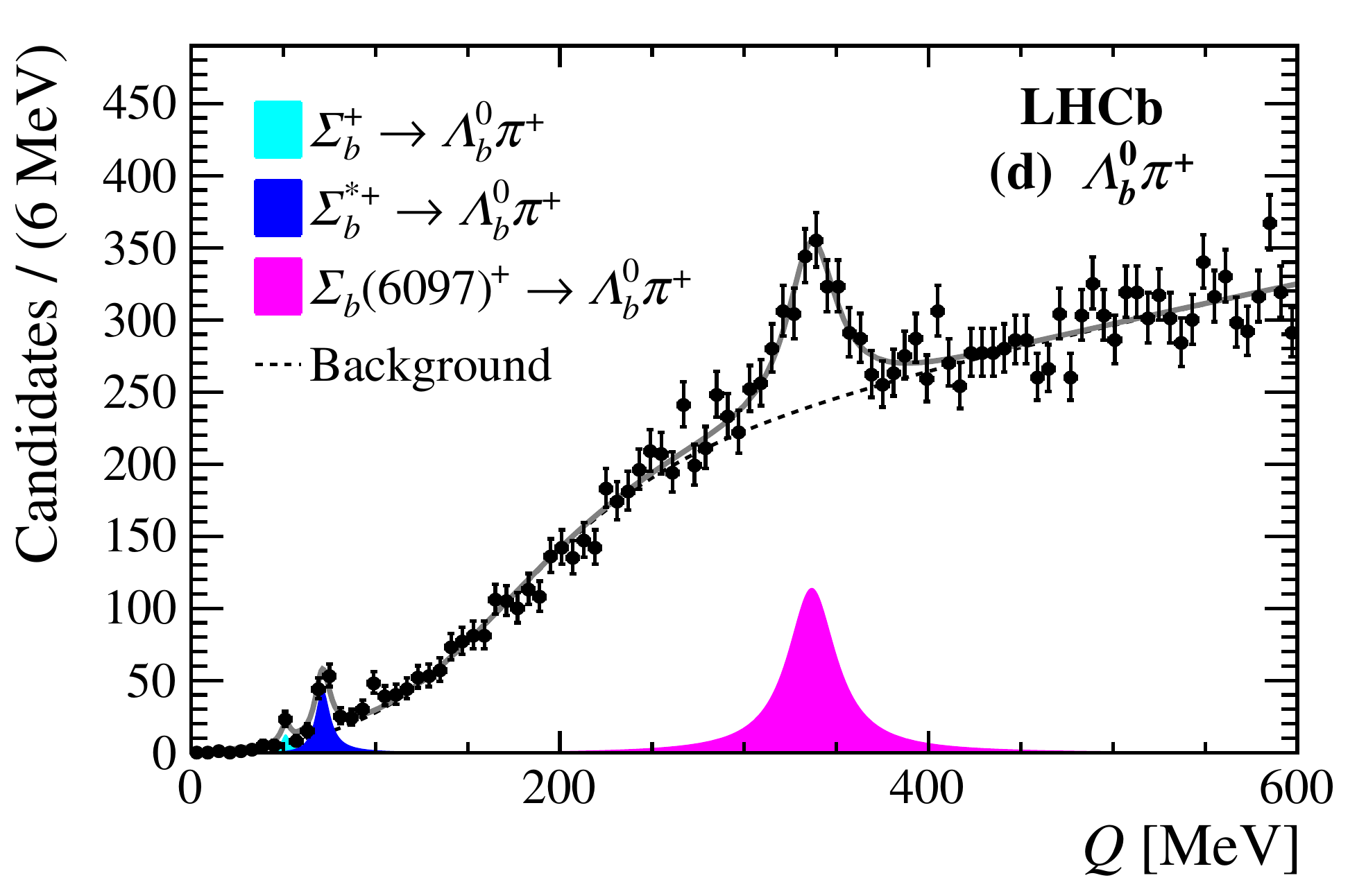}
\caption{\small
Mass distribution for selected $\lambdabpipm$ candidates. The points show experimental data.
The left (right) column shows $\lambdabpim$ ($\lambdabpip$) combinations.
The top row shows the fits to the lower-mass states $\sigmabpm$ and $\sigmabstarpm$.
The lower row presents the fits to the new mass peaks with the requirement $\pt(\pislow)>1000$\mev.
\label{fig:figure2}}
\end{figure}

The fits to the data sample are shown in Fig.~\ref{fig:figure2} and the resulting parameters of interest
are summarized in Table~\ref{tab:yields}.
The fit results are also used to determine mass differences and isospin splittings (given below).
The two new peaks in \lambdabpim and \lambdabpip distributions have local statistical significances of $12.7 \sigma$ and $12.6 \sigma$, respectively, based on the differences in log-likelihood between a fit with zero signal and the nominal fit.

\begin{table}[t]
\caption{\small Summary of the results of the fits to the \Qformula mass spectra.
$Q_0$ and $\Gamma$ are the mean and the width of the Breit--Wigner function.
The quoted uncertainties are statistical only.}
\centering
\vspace{2mm}
\begin{tabular}{lccc}
State        &  $Q_0$~$[\mevnospace]$        &  $\Gamma$~$[\mevnospace]$     &  Yield          \\
\hline
\sigmabm     & $\phantom{0}56.45 \pm 0.14$  &  $\phantom{0}5.33 \pm 0.42$  &  $3270 \pm 180$ \\
\sigmabstarm & $\phantom{0}75.54 \pm 0.17$  &  $10.68 \pm 0.60$            &  $7460 \pm 300$ \\
\sigmabp     & $\phantom{0}51.36 \pm 0.11$  &  $\phantom{0}4.83 \pm 0.31$  &  $3670 \pm 160$ \\
\sigmabstarp & $\phantom{0}71.09 \pm 0.14$  &  $\phantom{0}9.34 \pm 0.47$  &  $7350 \pm 260$ \\
\hline
\sigmabstarstarm &  $338.8 \pm 1.7$  &  $28.9 \pm 4.2$  &  $\phantom{0}880 \pm 100$ \\
\sigmabstarstarp &  $336.6 \pm 1.7$  &  $31.0 \pm 5.5$  &  $\phantom{0}900 \pm 110$ \\
\end{tabular}
\label{tab:yields}
\end{table}

\onecolumngrid

\twocolumngrid

%
%
%
%

Several sources of systematic uncertainty are considered.
The dominant source of systematic uncertainty on the mass measurements comes from the knowledge of the momentum scale.
This uncertainty is evaluated by adjusting the momentum scale by the $3 \times 10^{-4}$ relative uncertainty from the calibration procedure~\cite{LHCb-PAPER-2013-011} and rerunning the mass fit.
This procedure is also validated using a control sample of approximately 3 million $\Dstarp \rightarrow \Dz \pi^+$ decays,
with $\Dz \rightarrow K^- \pi^+$. 
The momentum-scale uncertainties largely cancel in the mass differences and splittings.
A second uncertainty arises from the parameterization of the background and is estimated by varying the function used
(\eg polynomial background functions of different order and other empirical curves).
An additional source of uncertainty on the determination of the natural widths arises from known differences in resolution between data and simulation. These are expected to agree within $5$\%, based on previous studies~\cite{LHCb-PAPER-2014-061,LHCb-PAPER-2017-002,LHCb-PAPER-2018-013},
and this assumption has been validated with the $\Dstarp \rightarrow \Dz \pi^+$ control sample.
Systematic uncertainties on the widths are assessed by varying the width of the resolution function by $\pm 5$\%.
Further uncertainties on the masses and widths arise from the assumed Breit--Wigner parameters.
The resonant states are assumed to decay to $\Lb \pi^\pm$ with angular momentum $l=1$.
For the \sigmabstarstarpm states, fits assuming $l=0,2,3$ are also performed and the largest changes to the fitted parameters with respect
to the nominal fit are assigned as systematic uncertainties.
The systematic uncertainties are summarised in Table~\ref{tab:sys}; in all cases they are much smaller than the statistical uncertainties.
All numerical results for the measured masses and widths are presented in Table~\ref{tab:resultstot}.
The mass values $m$ are obtained using the most precise \lhcb combination for the \Lb mass,
\mbox{$m(\Lb) = 5619.62 \pm 0.16 \pm 0.13$\mev}~\cite{LHCb-PAPER-2017-011},
which dominates by far the current world average~\cite{PDG2018}.
The correlated uncertainties, mainly deriving from the knowledge of the momentum scale
which is a common source of systematic uncertainty in all \lhcb mass measurements, 
are propagated as described in Ref.~\cite{LHCb-PAPER-2015-060}.
The isospin splittings of the new states are consistent with zero, although with large experimental uncertainty.

\begin{table}[t]
\caption{\small Summary of the systematic uncertainties on the measured masses and widths. 
$Q_0$ and $\Gamma$ are the mean and the width of the Breit--Wigner function.
All values are in \mev.}
\centering
\vspace{0.5mm}
\begin{tabular}{lccccccccc}
               & & \multicolumn{2}{c}{\sigmabm}& & \multicolumn{2}{c}{\sigmabstarm} & &\multicolumn{2}{c}{\sigmabstarstarm} \\
\hline
Source         & & $Q_0$ &  $\Gamma$           & & $Q_0$ &  $\Gamma$                & & $Q_0$ &  $\Gamma$                   \\
\hline
$p$ scale      & &	0.046 &	0.036 & &  0.047 & 0.071& &	0.130&	0.013\\
Resolution     & &	0.001 &	0.038 & &  0.001 & 0.033& &	0.003&	0.108\\
Spin assign.   & &	      &	      &	&        &	    & & 0.370&	0.462\\
Radius		   & &  0.003 & 0.101 & &  0.010 & 0.017& & 0.080&	0.081\\
Background	   & &	0.021 &	0.351 & &  0.033 & 0.315& & 0.184&	0.798\\
\hline
Total		   & &  0.051 & 0.369 &	&  0.058 & 0.325& & 0.440&	0.932\\
\end{tabular}

\vspace{3mm}

\begin{tabular}{lccccccccc}
               & & \multicolumn{2}{c}{\sigmabp}& & \multicolumn{2}{c}{\sigmabstarp} & &\multicolumn{2}{c}{\sigmabstarstarp} \\
\hline
Source         & & $Q_0$ &  $\Gamma$           & & $Q_0$ &  $\Gamma$                & & $Q_0$ &  $\Gamma$                   \\
\hline
$p$ scale      & & 0.039 &	0.046	& &	0.047	&	0.045	&	&	0.128	&	0.090	\\
Resolution	   & & 0.001 &	0.040	& &	0.001	&	0.038	&	&	0.002	&	0.086	\\
Spin assign.   & &		 &		    & &		    &		    &   &	0.113	&	0.342	\\
Radius		   & & 0.001 &	0.061	& &	0.003	&	0.002	&	&	0.001	&	0.031	\\
Background	   & & 0.027 &	0.357	& &	0.026	&	0.256	&	&	0.207	&	0.598	\\
\hline																				
Total		   & & 0.047 &	0.367	& &	0.053	&	0.263	&	&	0.268	&	0.701	\\
\end{tabular}
\label{tab:sys}
\end{table}
%
%
%
%
%

In summary, the first observation of two new mass peaks in the \lambdabpip and \lambdabpim systems is reported.
These structures are consistent with single resonances described by relativistic Breit--Wigner functions.
The ground-state \sigmabpm and \sigmabstarpm baryons are also confirmed and their masses and widths precisely measured.
These values are in good agreement with those measured by the \cdf collaboration~\cite{CDF:2011ac}, with precision improved by a factor of 5.
We also quote the mass differences and isospin splittings, for which most of the systematic uncertainties cancel.
\begin{table}[b]
\caption{\small Masses and widths of the \sigmabstarstarpm, \sigmabstarpm and \sigmabpm baryons.
Isospin splittings \mbox{$\Delta({X^\pm}) = m(X^+) - m(X^-)$} and mass differences are also calculated.
The first uncertainty is statistical, the second systematic.
The systematic uncertainty on $m$ includes the uncertainty from the knowledge of the \Lb mass~\cite{LHCb-PAPER-2017-011}.}
\centering
\vspace{0.5mm}
\begin{tabular}{r r @{$\,\pm\;$} c @{$\,\pm\;$} c }
\multicolumn{1}{c}{Quantity} &  \multicolumn{3}{c}{Value~$[\mevnospace]$}            \\
\hline
            $m(\sigmabstarstarm)$        & 6098.0  & 1.7  & 0.5 \\
            $m(\sigmabstarstarp)$        & 6095.8  & 1.7           & 0.4 \\
            $\Gamma(\sigmabstarstarm)$   & 28.9    & 4.2           & 0.9                 \\          
            $\Gamma(\sigmabstarstarp)$   & 31.0    & 5.5           & 0.7                 \\                                            
\hline
            $m(\sigmabm)$                & 5815.64  &  0.14  &  0.24 \\
            $m(\sigmabstarm)$            & 5834.73  &  0.17  &  0.25 \\
            $m(\sigmabp)$                & 5810.55  &  0.11  &  0.23 \\
            $m(\sigmabstarp)$            & 5830.28  &  0.14  &  0.24 \\
            $\Gamma(\sigmabm)$           & 5.33    &     0.42      &      0.37       \\
            $\Gamma(\sigmabstarm)$       & 10.68   &     0.60      &      0.33       \\
            $\Gamma(\sigmabp)$           & 4.83    &     0.31      &      0.37       \\
            $\Gamma(\sigmabstarp)$       & 9.34    &     0.47      &      0.26       \\
\hline
            $m(\sigmabstarm)-m(\sigmabm)$& 19.09   &    0.22       &     0.02          \\
            $m(\sigmabstarp)-m(\sigmabp)$& 19.73   &    0.18       &     0.01          \\
            $\Delta({\sigmabstarstarpm})$& $-2.2$  &      2.4     &     0.3          \\
            $\Delta({\sigmabpm})$        & $-5.09$ &      0.18     &     0.01          \\
            $\Delta({\sigmabstarpm})$    & $-4.45$ &      0.22     &     0.01          \\
\end{tabular}
\label{tab:resultstot}
\end{table}

In the heavy-quark limit, five $\sigmab(1P)$ states are expected.
Several predictions of their masses have been made~\cite{THAKKAR201757,PhysRevD.84.014025,PhysRevD.92.074026,newKarl},
but some or all of these states may be too wide to be accessible experimentally~\cite{PhysRevD.92.074026}.
Since the expected density of baryon states is high, it cannot be excluded that the new structures seen are the superpositions
of more than one (near-)degenerate state.
Taking into account that the predicted mass and width depend on the as-yet-unknown spin and parity,
the newly observed structures are compatible with being $1P$ excitations.
Other interpretations, such as molecular states, may also be possible~\cite{PhysRevD.89.054023}.


\section*{Acknowledgements}
%
%
\noindent  We thank Jonathan L. Rosner and Marek Karliner for useful discussions on the interpretation of the theoretical predictions.
We express our gratitude to our colleagues in the CERN
accelerator departments for the excellent performance of the LHC. We
thank the technical and administrative staff at the LHCb
institutes.
We acknowledge support from CERN and from the national agencies:
CAPES, CNPq, FAPERJ and FINEP (Brazil); 
MOST and NSFC (China); 
CNRS/IN2P3 (France); 
BMBF, DFG and MPG (Germany); 
INFN (Italy); 
NWO (Netherlands); 
MNiSW and NCN (Poland); 
MEN/IFA (Romania); 
MSHE (Russia); 
MinECo (Spain); 
SNSF and SER (Switzerland); 
NASU (Ukraine); 
STFC (United Kingdom); 
NSF (USA).
We acknowledge the computing resources that are provided by CERN, IN2P3
(France), KIT and DESY (Germany), INFN (Italy), SURF (Netherlands),
PIC (Spain), GridPP (United Kingdom), RRCKI and Yandex
LLC (Russia), CSCS (Switzerland), IFIN-HH (Romania), CBPF (Brazil),
PL-GRID (Poland) and OSC (USA).
We are indebted to the communities behind the multiple open-source
software packages on which we depend.
Individual groups or members have received support from
AvH Foundation (Germany);
EPLANET, Marie Sk\l{}odowska-Curie Actions and ERC (European Union);
ANR, Labex P2IO and OCEVU, and R\'{e}gion Auvergne-Rh\^{o}ne-Alpes (France);
Key Research Program of Frontier Sciences of CAS, CAS PIFI, and the Thousand Talents Program (China);
RFBR, RSF and Yandex LLC (Russia);
GVA, XuntaGal and GENCAT (Spain);
the Royal Society
and the Leverhulme Trust (United Kingdom);
Laboratory Directed Research and Development program of LANL (USA).



\clearpage
\addcontentsline{toc}{section}{References}
\setboolean{inbibliography}{true}
\bibliographystyle{LHCb}
\bibliography{standard,local,LHCb-PAPER,LHCb-CONF,LHCb-DP,LHCb-TDR}

\ifx\mcitethebibliography\mciteundefinedmacro
\PackageError{LHCb.bst}{mciteplus.sty has not been loaded}
{This bibstyle requires the use of the mciteplus package.}\fi
\providecommand{\href}[2]{#2}
\begin{mcitethebibliography}{10}
\mciteSetBstSublistMode{n}
\mciteSetBstMaxWidthForm{subitem}{\alph{mcitesubitemcount})}
\mciteSetBstSublistLabelBeginEnd{\mcitemaxwidthsubitemform\space}
{\relax}{\relax}

\bibitem{GELLMANN1964214}
M.~Gell-Mann, \ifthenelse{\boolean{articletitles}}{\emph{{A schematic model of
  baryons and mesons}},
  }{}\href{https://doi.org/10.1016/S0031-9163(64)92001-3}{Phys.\ Lett.\
  \textbf{8} (1964) 214}\relax
\mciteBstWouldAddEndPuncttrue
\mciteSetBstMidEndSepPunct{\mcitedefaultmidpunct}
{\mcitedefaultendpunct}{\mcitedefaultseppunct}\relax
\EndOfBibitem
\bibitem{Zweig:352337}
G.~Zweig, \ifthenelse{\boolean{articletitles}}{\emph{{An SU$_3$ model for
  strong interaction symmetry and its breaking; Version 1}}, }{} Tech. Rep.
  \href{http://cds.cern.ch/record/352337}{CERN-TH-401}, CERN, Geneva, Jan,
  1964\relax
\mciteBstWouldAddEndPuncttrue
\mciteSetBstMidEndSepPunct{\mcitedefaultmidpunct}
{\mcitedefaultendpunct}{\mcitedefaultseppunct}\relax
\EndOfBibitem
\bibitem{RevModPhys.82.1095}
E.~Klempt and J.-M. Richard, \ifthenelse{\boolean{articletitles}}{\emph{{Baryon
  spectroscopy}}, }{}\href{https://doi.org/10.1103/RevModPhys.82.1095}{Rev.\
  Mod.\ Phys.\  \textbf{82} (2010) 1095},
  \href{http://arxiv.org/abs/0901.2055}{{\normalfont\ttfamily
  arXiv:0901.2055}}\relax
\mciteBstWouldAddEndPuncttrue
\mciteSetBstMidEndSepPunct{\mcitedefaultmidpunct}
{\mcitedefaultendpunct}{\mcitedefaultseppunct}\relax
\EndOfBibitem
\bibitem{Aaltonen:2007ar}
CDF collaboration, T.~Aaltonen {\em et~al.},
  \ifthenelse{\boolean{articletitles}}{\emph{{Observation of the heavy baryons
  $\ensuremath{\Sigmares}_{b}$ and $\ensuremath{\Sigmares}_{b}^*$}},
  }{}\href{https://doi.org/10.1103/PhysRevLett.99.202001}{Phys.\ Rev.\ Lett.\
  \textbf{99} (2007) 202001},
  \href{http://arxiv.org/abs/0706.3868}{{\normalfont\ttfamily
  arXiv:0706.3868}}\relax
\mciteBstWouldAddEndPuncttrue
\mciteSetBstMidEndSepPunct{\mcitedefaultmidpunct}
{\mcitedefaultendpunct}{\mcitedefaultseppunct}\relax
\EndOfBibitem
\bibitem{CDF:2011ac}
CDF collaboration, T.~Aaltonen {\em et~al.},
  \ifthenelse{\boolean{articletitles}}{\emph{{Measurement of the masses and
  widths of the bottom baryons $\ensuremath{\Sigmares}_b^{\pm}$ and
  $\ensuremath{\Sigmares}_b^{*\pm}$}},
  }{}\href{https://doi.org/10.1103/PhysRevD.85.092011}{Phys.\ Rev.\
  \textbf{D85} (2012) 092011},
  \href{http://arxiv.org/abs/1112.2808}{{\normalfont\ttfamily
  arXiv:1112.2808}}\relax
\mciteBstWouldAddEndPuncttrue
\mciteSetBstMidEndSepPunct{\mcitedefaultmidpunct}
{\mcitedefaultendpunct}{\mcitedefaultseppunct}\relax
\EndOfBibitem
\bibitem{LHCb-PAPER-2015-047}
LHCb collaboration, R.~Aaij {\em et~al.},
  \ifthenelse{\boolean{articletitles}}{\emph{{Evidence for the
  strangeness-changing weak decay $\Xibm\to \Lb\pim$}},
  }{}\href{https://doi.org/10.1103/PhysRevLett.115.241801}{Phys.\ Rev.\ Lett.\
  \textbf{115} (2015) 241801},
  \href{http://arxiv.org/abs/1510.03829}{{\normalfont\ttfamily
  arXiv:1510.03829}}\relax
\mciteBstWouldAddEndPuncttrue
\mciteSetBstMidEndSepPunct{\mcitedefaultmidpunct}
{\mcitedefaultendpunct}{\mcitedefaultseppunct}\relax
\EndOfBibitem
\bibitem{LHCb-PAPER-2012-012}
LHCb collaboration, R.~Aaij {\em et~al.},
  \ifthenelse{\boolean{articletitles}}{\emph{{Observation of excited $\Lb$
  baryons}}, }{}\href{https://doi.org/10.1103/PhysRevLett.109.172003}{Phys.\
  Rev.\ Lett.\  \textbf{109} (2012) 172003},
  \href{http://arxiv.org/abs/1205.3452}{{\normalfont\ttfamily
  arXiv:1205.3452}}\relax
\mciteBstWouldAddEndPuncttrue
\mciteSetBstMidEndSepPunct{\mcitedefaultmidpunct}
{\mcitedefaultendpunct}{\mcitedefaultseppunct}\relax
\EndOfBibitem
\bibitem{LHCb-PAPER-2018-013}
LHCb collaboration, R.~Aaij {\em et~al.},
  \ifthenelse{\boolean{articletitles}}{\emph{{Observation of a new
  $\Xires_b^{-}$ resonance}},
  }{}\href{https://doi.org/10.1103/PhysRevLett.121.072002}{Phys.\ Rev.\ Lett.\
  \textbf{121} (2018) 072002},
  \href{http://arxiv.org/abs/1805.09418}{{\normalfont\ttfamily
  arXiv:1805.09418}}\relax
\mciteBstWouldAddEndPuncttrue
\mciteSetBstMidEndSepPunct{\mcitedefaultmidpunct}
{\mcitedefaultendpunct}{\mcitedefaultseppunct}\relax
\EndOfBibitem
\bibitem{PhysRevLett.108.252002}
CMS collaboration, S.~Chatrchyan {\em et~al.},
  \ifthenelse{\boolean{articletitles}}{\emph{{Observation of a new
  ${\ensuremath{\Xires}}_{b}$ baryon}},
  }{}\href{https://doi.org/10.1103/PhysRevLett.108.252002}{Phys.\ Rev.\ Lett.\
  \textbf{108} (2012) 252002},
  \href{http://arxiv.org/abs/1204.5955}{{\normalfont\ttfamily
  arXiv:1204.5955}}\relax
\mciteBstWouldAddEndPuncttrue
\mciteSetBstMidEndSepPunct{\mcitedefaultmidpunct}
{\mcitedefaultendpunct}{\mcitedefaultseppunct}\relax
\EndOfBibitem
\bibitem{LHCb-PAPER-2014-061}
LHCb collaboration, R.~Aaij {\em et~al.},
  \ifthenelse{\boolean{articletitles}}{\emph{{Observation of two new $\Xibm$
  baryon resonances}},
  }{}\href{https://doi.org/10.1103/PhysRevLett.114.062004}{Phys.\ Rev.\ Lett.\
  \textbf{114} (2015) 062004},
  \href{http://arxiv.org/abs/1411.4849}{{\normalfont\ttfamily
  arXiv:1411.4849}}\relax
\mciteBstWouldAddEndPuncttrue
\mciteSetBstMidEndSepPunct{\mcitedefaultmidpunct}
{\mcitedefaultendpunct}{\mcitedefaultseppunct}\relax
\EndOfBibitem
\bibitem{THAKKAR201757}
K.~Thakkar, Z.~Shah, A.~K. Rai, and P.~C. Vinodkumar,
  \ifthenelse{\boolean{articletitles}}{\emph{{Excited state mass spectra and
  Regge trajectories of bottom baryons}},
  }{}\href{https://doi.org/10.1016/j.nuclphysa.2017.05.087}{Nucl.\ Phys.\
  \textbf{A965} (2017) 57},
  \href{http://arxiv.org/abs/1610.00411}{{\normalfont\ttfamily
  arXiv:1610.00411}}\relax
\mciteBstWouldAddEndPuncttrue
\mciteSetBstMidEndSepPunct{\mcitedefaultmidpunct}
{\mcitedefaultendpunct}{\mcitedefaultseppunct}\relax
\EndOfBibitem
\bibitem{PhysRevD.84.014025}
D.~Ebert, R.~N. Faustov, and V.~O. Galkin,
  \ifthenelse{\boolean{articletitles}}{\emph{{Spectroscopy and Regge
  trajectories of heavy baryons in the relativistic quark-diquark picture}},
  }{}\href{https://doi.org/10.1103/PhysRevD.84.014025}{Phys.\ Rev.\
  \textbf{D84} (2011) 014025},
  \href{http://arxiv.org/abs/1105.0583}{{\normalfont\ttfamily
  arXiv:1105.0583}}\relax
\mciteBstWouldAddEndPuncttrue
\mciteSetBstMidEndSepPunct{\mcitedefaultmidpunct}
{\mcitedefaultendpunct}{\mcitedefaultseppunct}\relax
\EndOfBibitem
\bibitem{Alves:2008zz}
LHCb collaboration, A.~A. Alves~Jr.\ {\em et~al.},
  \ifthenelse{\boolean{articletitles}}{\emph{{The \lhcb detector at the LHC}},
  }{}\href{https://doi.org/10.1088/1748-0221/3/08/S08005}{JINST \textbf{3}
  (2008) S08005}\relax
\mciteBstWouldAddEndPuncttrue
\mciteSetBstMidEndSepPunct{\mcitedefaultmidpunct}
{\mcitedefaultendpunct}{\mcitedefaultseppunct}\relax
\EndOfBibitem
\bibitem{LHCb-DP-2014-002}
LHCb collaboration, R.~Aaij {\em et~al.},
  \ifthenelse{\boolean{articletitles}}{\emph{{LHCb detector performance}},
  }{}\href{https://doi.org/10.1142/S0217751X15300227}{Int.\ J.\ Mod.\ Phys.\
  \textbf{A30} (2015) 1530022},
  \href{http://arxiv.org/abs/1412.6352}{{\normalfont\ttfamily
  arXiv:1412.6352}}\relax
\mciteBstWouldAddEndPuncttrue
\mciteSetBstMidEndSepPunct{\mcitedefaultmidpunct}
{\mcitedefaultendpunct}{\mcitedefaultseppunct}\relax
\EndOfBibitem
\bibitem{LHCb-DP-2014-001}
R.~Aaij {\em et~al.}, \ifthenelse{\boolean{articletitles}}{\emph{{Performance
  of the LHCb Vertex Locator}},
  }{}\href{https://doi.org/10.1088/1748-0221/9/09/P09007}{JINST \textbf{9}
  (2014) P09007}, \href{http://arxiv.org/abs/1405.7808}{{\normalfont\ttfamily
  arXiv:1405.7808}}\relax
\mciteBstWouldAddEndPuncttrue
\mciteSetBstMidEndSepPunct{\mcitedefaultmidpunct}
{\mcitedefaultendpunct}{\mcitedefaultseppunct}\relax
\EndOfBibitem
\bibitem{LHCb-DP-2013-003}
R.~Arink {\em et~al.}, \ifthenelse{\boolean{articletitles}}{\emph{{Performance
  of the LHCb Outer Tracker}},
  }{}\href{https://doi.org/10.1088/1748-0221/9/01/P01002}{JINST \textbf{9}
  (2014) P01002}, \href{http://arxiv.org/abs/1311.3893}{{\normalfont\ttfamily
  arXiv:1311.3893}}\relax
\mciteBstWouldAddEndPuncttrue
\mciteSetBstMidEndSepPunct{\mcitedefaultmidpunct}
{\mcitedefaultendpunct}{\mcitedefaultseppunct}\relax
\EndOfBibitem
\bibitem{LHCb-PAPER-2012-048}
LHCb collaboration, R.~Aaij {\em et~al.},
  \ifthenelse{\boolean{articletitles}}{\emph{{Measurements of the $\Lb$,
  $\Xibm$, and $\Omegab$ baryon masses}},
  }{}\href{https://doi.org/10.1103/PhysRevLett.110.182001}{Phys.\ Rev.\ Lett.\
  \textbf{110} (2013) 182001},
  \href{http://arxiv.org/abs/1302.1072}{{\normalfont\ttfamily
  arXiv:1302.1072}}\relax
\mciteBstWouldAddEndPuncttrue
\mciteSetBstMidEndSepPunct{\mcitedefaultmidpunct}
{\mcitedefaultendpunct}{\mcitedefaultseppunct}\relax
\EndOfBibitem
\bibitem{LHCb-PAPER-2013-011}
LHCb collaboration, R.~Aaij {\em et~al.},
  \ifthenelse{\boolean{articletitles}}{\emph{{Precision measurement of $\D$
  meson mass differences}},
  }{}\href{https://doi.org/10.1007/JHEP06(2013)065}{JHEP \textbf{06} (2013)
  065}, \href{http://arxiv.org/abs/1304.6865}{{\normalfont\ttfamily
  arXiv:1304.6865}}\relax
\mciteBstWouldAddEndPuncttrue
\mciteSetBstMidEndSepPunct{\mcitedefaultmidpunct}
{\mcitedefaultendpunct}{\mcitedefaultseppunct}\relax
\EndOfBibitem
\bibitem{LHCb-DP-2012-003}
M.~Adinolfi {\em et~al.},
  \ifthenelse{\boolean{articletitles}}{\emph{{Performance of the \lhcb RICH
  detector at the LHC}},
  }{}\href{https://doi.org/10.1140/epjc/s10052-013-2431-9}{Eur.\ Phys.\ J.\
  \textbf{C73} (2013) 2431},
  \href{http://arxiv.org/abs/1211.6759}{{\normalfont\ttfamily
  arXiv:1211.6759}}\relax
\mciteBstWouldAddEndPuncttrue
\mciteSetBstMidEndSepPunct{\mcitedefaultmidpunct}
{\mcitedefaultendpunct}{\mcitedefaultseppunct}\relax
\EndOfBibitem
\bibitem{LHCb-DP-2012-004}
R.~Aaij {\em et~al.}, \ifthenelse{\boolean{articletitles}}{\emph{{The \lhcb
  trigger and its performance in 2011}},
  }{}\href{https://doi.org/10.1088/1748-0221/8/04/P04022}{JINST \textbf{8}
  (2013) P04022}, \href{http://arxiv.org/abs/1211.3055}{{\normalfont\ttfamily
  arXiv:1211.3055}}\relax
\mciteBstWouldAddEndPuncttrue
\mciteSetBstMidEndSepPunct{\mcitedefaultmidpunct}
{\mcitedefaultendpunct}{\mcitedefaultseppunct}\relax
\EndOfBibitem
\bibitem{BBDT}
V.~V. Gligorov and M.~Williams,
  \ifthenelse{\boolean{articletitles}}{\emph{{Efficient, reliable and fast
  high-level triggering using a bonsai boosted decision tree}},
  }{}\href{https://doi.org/10.1088/1748-0221/8/02/P02013}{JINST \textbf{8}
  (2013) P02013}, \href{http://arxiv.org/abs/1210.6861}{{\normalfont\ttfamily
  arXiv:1210.6861}}\relax
\mciteBstWouldAddEndPuncttrue
\mciteSetBstMidEndSepPunct{\mcitedefaultmidpunct}
{\mcitedefaultendpunct}{\mcitedefaultseppunct}\relax
\EndOfBibitem
\bibitem{Sjostrand:2006za}
T.~Sj\"{o}strand, S.~Mrenna, and P.~Skands,
  \ifthenelse{\boolean{articletitles}}{\emph{{PYTHIA 6.4 physics and manual}},
  }{}\href{https://doi.org/10.1088/1126-6708/2006/05/026}{JHEP \textbf{05}
  (2006) 026}, \href{http://arxiv.org/abs/hep-ph/0603175}{{\normalfont\ttfamily
  arXiv:hep-ph/0603175}}\relax
\mciteBstWouldAddEndPuncttrue
\mciteSetBstMidEndSepPunct{\mcitedefaultmidpunct}
{\mcitedefaultendpunct}{\mcitedefaultseppunct}\relax
\EndOfBibitem
\bibitem{Sjostrand:2007gs}
T.~Sj\"{o}strand, S.~Mrenna, and P.~Skands,
  \ifthenelse{\boolean{articletitles}}{\emph{{A brief introduction to PYTHIA
  8.1}}, }{}\href{https://doi.org/10.1016/j.cpc.2008.01.036}{Comput.\ Phys.\
  Commun.\  \textbf{178} (2008) 852},
  \href{http://arxiv.org/abs/0710.3820}{{\normalfont\ttfamily
  arXiv:0710.3820}}\relax
\mciteBstWouldAddEndPuncttrue
\mciteSetBstMidEndSepPunct{\mcitedefaultmidpunct}
{\mcitedefaultendpunct}{\mcitedefaultseppunct}\relax
\EndOfBibitem
\bibitem{LHCb-PROC-2010-056}
I.~Belyaev {\em et~al.}, \ifthenelse{\boolean{articletitles}}{\emph{{Handling
  of the generation of primary events in Gauss, the LHCb simulation
  framework}}, }{}\href{https://doi.org/10.1088/1742-6596/331/3/032047}{J.\
  Phys.\ Conf.\ Ser.\  \textbf{331} (2011) 032047}\relax
\mciteBstWouldAddEndPuncttrue
\mciteSetBstMidEndSepPunct{\mcitedefaultmidpunct}
{\mcitedefaultendpunct}{\mcitedefaultseppunct}\relax
\EndOfBibitem
\bibitem{Lange:2001uf}
D.~J. Lange, \ifthenelse{\boolean{articletitles}}{\emph{{The EvtGen particle
  decay simulation package}},
  }{}\href{https://doi.org/10.1016/S0168-9002(01)00089-4}{Nucl.\ Instrum.\
  Meth.\  \textbf{A462} (2001) 152}\relax
\mciteBstWouldAddEndPuncttrue
\mciteSetBstMidEndSepPunct{\mcitedefaultmidpunct}
{\mcitedefaultendpunct}{\mcitedefaultseppunct}\relax
\EndOfBibitem
\bibitem{Golonka:2005pn}
P.~Golonka and Z.~Was, \ifthenelse{\boolean{articletitles}}{\emph{{PHOTOS Monte
  Carlo: A precision tool for QED corrections in $Z$ and $W$ decays}},
  }{}\href{https://doi.org/10.1140/epjc/s2005-02396-4}{Eur.\ Phys.\ J.\
  \textbf{C45} (2006) 97},
  \href{http://arxiv.org/abs/hep-ph/0506026}{{\normalfont\ttfamily
  arXiv:hep-ph/0506026}}\relax
\mciteBstWouldAddEndPuncttrue
\mciteSetBstMidEndSepPunct{\mcitedefaultmidpunct}
{\mcitedefaultendpunct}{\mcitedefaultseppunct}\relax
\EndOfBibitem
\bibitem{Allison:2006ve}
Geant4 collaboration, J.~Allison {\em et~al.},
  \ifthenelse{\boolean{articletitles}}{\emph{{Geant4 developments and
  applications}}, }{}\href{https://doi.org/10.1109/TNS.2006.869826}{IEEE
  Trans.\ Nucl.\ Sci.\  \textbf{53} (2006) 270}\relax
\mciteBstWouldAddEndPuncttrue
\mciteSetBstMidEndSepPunct{\mcitedefaultmidpunct}
{\mcitedefaultendpunct}{\mcitedefaultseppunct}\relax
\EndOfBibitem
\bibitem{Agostinelli:2002hh}
Geant4 collaboration, S.~Agostinelli {\em et~al.},
  \ifthenelse{\boolean{articletitles}}{\emph{{Geant4: A simulation toolkit}},
  }{}\href{https://doi.org/10.1016/S0168-9002(03)01368-8}{Nucl.\ Instrum.\
  Meth.\  \textbf{A506} (2003) 250}\relax
\mciteBstWouldAddEndPuncttrue
\mciteSetBstMidEndSepPunct{\mcitedefaultmidpunct}
{\mcitedefaultendpunct}{\mcitedefaultseppunct}\relax
\EndOfBibitem
\bibitem{DeCian:2255039}
M.~De~Cian, S.~Farry, P.~Seyfert, and S.~Stahl,
  \ifthenelse{\boolean{articletitles}}{\emph{{Fast neural-net based fake track
  rejection in the LHCb reconstruction}}, }{}
  \href{http://cdsweb.cern.ch/search?p=LHCb-PUB-2017-011&f=reportnumber&action_search=Search&c=LHCb+Notes}
  {LHCb-PUB-2017-011}\relax
\mciteBstWouldAddEndPuncttrue
\mciteSetBstMidEndSepPunct{\mcitedefaultmidpunct}
{\mcitedefaultendpunct}{\mcitedefaultseppunct}\relax
\EndOfBibitem
\bibitem{PDG2016}
Particle Data Group, C.~Patrignani {\em et~al.},
  \ifthenelse{\boolean{articletitles}}{\emph{{\href{http://pdg.lbl.gov/}{Review
  of particle physics}}},
  }{}\href{https://doi.org/10.1088/1674-1137/40/10/100001}{Chin.\ Phys.\
  \textbf{C40} (2016) 100001}\relax
\mciteBstWouldAddEndPuncttrue
\mciteSetBstMidEndSepPunct{\mcitedefaultmidpunct}
{\mcitedefaultendpunct}{\mcitedefaultseppunct}\relax
\EndOfBibitem
\bibitem{ROE2005577}
B.~P. Roe {\em et~al.}, \ifthenelse{\boolean{articletitles}}{\emph{{Boosted
  decision trees as an alternative to artificial neural networks for particle
  identification}}, }{}\href{https://doi.org/10.1016/j.nima.2004.12.018}{Nucl.\
  Instrum.\ Meth.\  \textbf{A543} (2005) 577}\relax
\mciteBstWouldAddEndPuncttrue
\mciteSetBstMidEndSepPunct{\mcitedefaultmidpunct}
{\mcitedefaultendpunct}{\mcitedefaultseppunct}\relax
\EndOfBibitem
\bibitem{AdaBoost}
Y.~Freund and R.~E. Schapire, \ifthenelse{\boolean{articletitles}}{\emph{A
  decision-theoretic generalization of on-line learning and an application to
  boosting}, }{}\href{https://doi.org/10.1006/jcss.1997.1504}{J.\ Comput.\
  Syst.\ Sci.\  \textbf{55} (1997) 119}\relax
\mciteBstWouldAddEndPuncttrue
\mciteSetBstMidEndSepPunct{\mcitedefaultmidpunct}
{\mcitedefaultendpunct}{\mcitedefaultseppunct}\relax
\EndOfBibitem
\bibitem{Hulsbergen:2005pu}
W.~D. Hulsbergen, \ifthenelse{\boolean{articletitles}}{\emph{{Decay chain
  fitting with a Kalman filter}},
  }{}\href{https://doi.org/10.1016/j.nima.2005.06.078}{Nucl.\ Instrum.\ Meth.\
  \textbf{A552} (2005) 566},
  \href{http://arxiv.org/abs/physics/0503191}{{\normalfont\ttfamily
  arXiv:physics/0503191}}\relax
\mciteBstWouldAddEndPuncttrue
\mciteSetBstMidEndSepPunct{\mcitedefaultmidpunct}
{\mcitedefaultendpunct}{\mcitedefaultseppunct}\relax
\EndOfBibitem
\bibitem{LHCb-PAPER-2017-021}
LHCb collaboration, R.~Aaij {\em et~al.},
  \ifthenelse{\boolean{articletitles}}{\emph{{Measurement of \CP observables in
  $\Bpm\to D^{(\ast)}K^\pm$ and $\Bpm\to D^{(\ast)}\pi^\pm$ decays}},
  }{}\href{https://doi.org/10.1016/j.physletb.2017.11.070}{Phys.\ Lett.\
  \textbf{B777} (2017) 16},
  \href{http://arxiv.org/abs/1708.06370}{{\normalfont\ttfamily
  arXiv:1708.06370}}\relax
\mciteBstWouldAddEndPuncttrue
\mciteSetBstMidEndSepPunct{\mcitedefaultmidpunct}
{\mcitedefaultendpunct}{\mcitedefaultseppunct}\relax
\EndOfBibitem
\bibitem{Jackson1964}
J.~D. Jackson, \ifthenelse{\boolean{articletitles}}{\emph{{Remarks on the
  phenomenological analysis of resonances}},
  }{}\href{https://doi.org/10.1007/BF02750563}{Il Nuovo Cimento \textbf{34}
  (1964) 1644}\relax
\mciteBstWouldAddEndPuncttrue
\mciteSetBstMidEndSepPunct{\mcitedefaultmidpunct}
{\mcitedefaultendpunct}{\mcitedefaultseppunct}\relax
\EndOfBibitem
\bibitem{Blatt:1952ije}
J.~M. Blatt and V.~F. Weisskopf, {\em {Theoretical nuclear physics}},
  \href{https://doi.org/10.1007/978-1-4612-9959-2}{ Springer, New York,
  1952}\relax
\mciteBstWouldAddEndPuncttrue
\mciteSetBstMidEndSepPunct{\mcitedefaultmidpunct}
{\mcitedefaultendpunct}{\mcitedefaultseppunct}\relax
\EndOfBibitem
\bibitem{LHCb-PAPER-2012-030}
LHCb collaboration, R.~Aaij {\em et~al.},
  \ifthenelse{\boolean{articletitles}}{\emph{{First observation of the decay
  $B_{s2}^*(5840)^0 \to B^{*+}\Km$ and studies of excited $\Bs$ mesons}},
  }{}\href{https://doi.org/10.1103/PhysRevLett.110.151803}{Phys.\ Rev.\ Lett.\
  \textbf{110} (2013) 151803},
  \href{http://arxiv.org/abs/1211.5994}{{\normalfont\ttfamily
  arXiv:1211.5994}}\relax
\mciteBstWouldAddEndPuncttrue
\mciteSetBstMidEndSepPunct{\mcitedefaultmidpunct}
{\mcitedefaultendpunct}{\mcitedefaultseppunct}\relax
\EndOfBibitem
\bibitem{LHCb-PAPER-2017-002}
LHCb collaboration, R.~Aaij {\em et~al.},
  \ifthenelse{\boolean{articletitles}}{\emph{{Observation of five new narrow
  $\Omegac$ states decaying to $\Xicp\Km$}},
  }{}\href{https://doi.org/10.1103/PhysRevLett.118.182001}{Phys.\ Rev.\ Lett.\
  \textbf{118} (2017) 182001},
  \href{http://arxiv.org/abs/1703.04639}{{\normalfont\ttfamily
  arXiv:1703.04639}}\relax
\mciteBstWouldAddEndPuncttrue
\mciteSetBstMidEndSepPunct{\mcitedefaultmidpunct}
{\mcitedefaultendpunct}{\mcitedefaultseppunct}\relax
\EndOfBibitem
\bibitem{LHCb-PAPER-2013-028}
LHCb collaboration, R.~Aaij {\em et~al.},
  \ifthenelse{\boolean{articletitles}}{\emph{{Measurement of the relative rate
  of prompt $\chiczero$, $\chicone$ and $\chictwo$ production at
  $\sqrt{s}=7$\tev}}, }{}\href{https://doi.org/10.1007/JHEP10(2013)115}{JHEP
  \textbf{10} (2013) 115},
  \href{http://arxiv.org/abs/1307.4285}{{\normalfont\ttfamily
  arXiv:1307.4285}}\relax
\mciteBstWouldAddEndPuncttrue
\mciteSetBstMidEndSepPunct{\mcitedefaultmidpunct}
{\mcitedefaultendpunct}{\mcitedefaultseppunct}\relax
\EndOfBibitem
\bibitem{LHCb-PAPER-2011-019}
LHCb collaboration, R.~Aaij {\em et~al.},
  \ifthenelse{\boolean{articletitles}}{\emph{{Measurement of the cross-section
  ratio $\sigma(\chictwo)/\sigma(\chicone)$ for prompt $\chi_c$ production at
  $\sqrt{s}=7$\tev}},
  }{}\href{https://doi.org/10.1016/j.physletb.2012.06.077}{Phys.\ Lett.\
  \textbf{B714} (2012) 215},
  \href{http://arxiv.org/abs/1202.1080}{{\normalfont\ttfamily
  arXiv:1202.1080}}\relax
\mciteBstWouldAddEndPuncttrue
\mciteSetBstMidEndSepPunct{\mcitedefaultmidpunct}
{\mcitedefaultendpunct}{\mcitedefaultseppunct}\relax
\EndOfBibitem
\bibitem{LHCb-PAPER-2017-011}
LHCb collaboration, R.~Aaij {\em et~al.},
  \ifthenelse{\boolean{articletitles}}{\emph{{Observation of the decays $\Lb\to
  \chicone\proton\Km$ and $\Lb\to \chictwo\proton\Km$}},
  }{}\href{https://doi.org/10.1103/PhysRevLett.119.062001}{Phys.\ Rev.\ Lett.\
  \textbf{119} (2017) 062001},
  \href{http://arxiv.org/abs/1704.07900}{{\normalfont\ttfamily
  arXiv:1704.07900}}\relax
\mciteBstWouldAddEndPuncttrue
\mciteSetBstMidEndSepPunct{\mcitedefaultmidpunct}
{\mcitedefaultendpunct}{\mcitedefaultseppunct}\relax
\EndOfBibitem
\bibitem{PDG2018}
Particle Data Group, M.~Tanabashi {\em et~al.},
  \ifthenelse{\boolean{articletitles}}{\emph{{\href{http://pdg.lbl.gov/}{Review
  of particle physics}}},
  }{}\href{https://doi.org/10.1103/PhysRevD.98.030001}{Phys.\ Rev.\
  \textbf{D98} (2018) 030001}\relax
\mciteBstWouldAddEndPuncttrue
\mciteSetBstMidEndSepPunct{\mcitedefaultmidpunct}
{\mcitedefaultendpunct}{\mcitedefaultseppunct}\relax
\EndOfBibitem
\bibitem{LHCb-PAPER-2015-060}
LHCb collaboration, R.~Aaij {\em et~al.},
  \ifthenelse{\boolean{articletitles}}{\emph{{Observation of $\Lb\to
  \psitwos\proton\Km$ and $\Lb\to \jpsi\pip\pim\proton\Km$ decays and a
  measurement of the $\Lb$ baryon mass}},
  }{}\href{https://doi.org/10.1007/JHEP05(2016)132}{JHEP \textbf{05} (2016)
  132}, \href{http://arxiv.org/abs/1603.06961}{{\normalfont\ttfamily
  arXiv:1603.06961}}\relax
\mciteBstWouldAddEndPuncttrue
\mciteSetBstMidEndSepPunct{\mcitedefaultmidpunct}
{\mcitedefaultendpunct}{\mcitedefaultseppunct}\relax
\EndOfBibitem
\bibitem{PhysRevD.92.074026}
M.~Karliner and J.~L. Rosner,
  \ifthenelse{\boolean{articletitles}}{\emph{{Prospects for observing the
  lowest-lying odd-parity ${\Sigmares}_{c}$ and ${\Sigmares}_{b}$ baryons}},
  }{}\href{https://doi.org/10.1103/PhysRevD.92.074026}{Phys.\ Rev.\
  \textbf{D92} (2015) 074026},
  \href{http://arxiv.org/abs/1506.01702}{{\normalfont\ttfamily
  arXiv:1506.01702}}\relax
\mciteBstWouldAddEndPuncttrue
\mciteSetBstMidEndSepPunct{\mcitedefaultmidpunct}
{\mcitedefaultendpunct}{\mcitedefaultseppunct}\relax
\EndOfBibitem
\bibitem{newKarl}
M.~Karliner and J.~L. Rosner,
  \ifthenelse{\boolean{articletitles}}{\emph{{Scaling of P-wave excitation
  energies in heavy-quark systems}},
  }{}\href{http://arxiv.org/abs/1808.07869}{{\normalfont\ttfamily
  arXiv:1808.07869}}\relax
\mciteBstWouldAddEndPuncttrue
\mciteSetBstMidEndSepPunct{\mcitedefaultmidpunct}
{\mcitedefaultendpunct}{\mcitedefaultseppunct}\relax
\EndOfBibitem
\bibitem{PhysRevD.89.054023}
W.~H. Liang, C.~W. Xiao, and E.~Oset,
  \ifthenelse{\boolean{articletitles}}{\emph{{Baryon states with open beauty in
  the extended local hidden gauge approach}},
  }{}\href{https://doi.org/10.1103/PhysRevD.89.054023}{Phys.\ Rev.\
  \textbf{D89} (2014) 054023}\relax
\mciteBstWouldAddEndPuncttrue
\mciteSetBstMidEndSepPunct{\mcitedefaultmidpunct}
{\mcitedefaultendpunct}{\mcitedefaultseppunct}\relax
\EndOfBibitem
\end{mcitethebibliography}

\newpage

\centerline{\large\bf LHCb collaboration}
\begin{flushleft}
\small
R.~Aaij$^{27}$,
C.~Abell{\'a}n~Beteta$^{44}$,
B.~Adeva$^{41}$,
M.~Adinolfi$^{48}$,
C.A.~Aidala$^{74}$,
Z.~Ajaltouni$^{5}$,
S.~Akar$^{59}$,
P.~Albicocco$^{18}$,
J.~Albrecht$^{10}$,
F.~Alessio$^{42}$
% ...

{\footnotesize \it
% ...
$ ^{5}$Clermont Universit{\'e}, Universit{\'e} Blaise Pascal, CNRS/IN2P3, LPC, Clermont-Ferrand, France\\
% ...
$ ^{10}$Fakult{\"a}t Physik, Technische Universit{\"a}t Dortmund, Dortmund, Germany\\
% ...
$ ^{18}$INFN Laboratori Nazionali di Frascati, Frascati, Italy\\
% ...
$ ^{27}$Nikhef National Institute for Subatomic Physics, Amsterdam, Netherlands\\
% ...
$ ^{41}$Instituto Galego de F{\'\i}sica de Altas Enerx{\'\i}as (IGFAE), Universidade de Santiago de Compostela, Santiago de Compostela, Spain\\
$ ^{42}$European Organization for Nuclear Research (CERN), Geneva, Switzerland\\
% ...
$ ^{44}$Physik-Institut, Universit{\"a}t Z{\"u}rich, Z{\"u}rich, Switzerland\\
% ...
$ ^{48}$H.H. Wills Physics Laboratory, University of Bristol, Bristol, United Kingdom\\
% ...
$ ^{59}$University of Cincinnati, Cincinnati, OH, United States\\
% ...
$ ^{61}$Syracuse University, Syracuse, NY, United States\\
% ...
$ ^{74}$University of Michigan, Ann Arbor, United States, associated to $^{61}$\\
% ...
}
\end{flushleft}

\end{document}